\documentclass{JHEP3}
\usepackage{amssymb}

\def \be {\begin{equation}}
\def \ee {\end{equation}}

\def \bea {\begin{eqnarray}}
\def \eea {\end{eqnarray}}
\def \nn {\nonumber}

\def \a {\alpha}
\def \b {\beta}

\def \d {\delta}

\def \m {\mu}
\def \n {\nu}
\def \k {\kappa}

\def \s {\sigma}
\def \r {\rho}
\def \o {\omega}

\def \th {\theta}
\def \Th {\Theta}

\def \t {\tau}
\def \dag {\dagger}
\def \p {\partial}

\def\bd{\begin{document}}
\def\ed{\end{document}}
\def\nn{\nonumber}
\def\bea{\begin{eqnarray}}
\def\eea{\end{eqnarray}}
\let\bm=\bibitem
\let\la=\label

\def\N{{\cal N}}
\def\sst{\scriptscriptstyle}
\def\thetabar{\bar\theta}
\def\Tr{{\rm Tr}}
\def\one{\mbox{1 \kern-.59em {\rm l}}}

\def\a{\alpha}      \def\da{{\dot\alpha}}
\def\b{\beta}       \def\db{{\dot\beta}}
\def\c{\gamma}  \def\C{\Gamma}  \def\cdt{\dot\gamma}
\def\d{\delta}  \def\D{\Delta}  \def\ddt{\dot\delta}
\def\e{\epsilon}        \def\vare{\varepsilon}
\def\f{\phi}    \def\F{\Phi}    \def\vvf{\f}
\def\h{\eta}
\def\k{\kappa}
\def\l{\lambda} \def\L{\Lambda}
\def\m{\mu} \def\n{\nu}
\def\o{\omega}
\def\P{\Pi}
\def\r{\rho}
\def\s{\sigma}  \def\S{\Sigma}
\def\t{\tau}
\def\th{\theta} \def\Th{\Theta} \def\vth{\vartheta}
\def\X{\Xeta}
\def\z{\zeta}
\def\w{\wedge}
\def\u{\underline}
\def\hs{\hspace}

\def\cA{{\cal A}} \def\cB{{\cal B}} \def\cC{{\cal C}}
\def\cD{{\cal D}} \def\cE{{\cal E}} \def\cF{{\cal F}}
\def\cG{{\cal G}} \def\cH{{\cal H}} \def\cI{{\cal I}}
\def\cJ{{\cal J}} \def\cK{{\cal K}} \def\cL{{\cal L}}
\def\cM{{\cal M}} \def\cN{{\cal N}} \def\cO{{\cal O}}
\def\cP{{\cal P}} \def\cQ{{\cal Q}} \def\cR{{\cal R}}
\def\cS{{\cal S}} \def\cT{{\cal T}} \def\cU{{\cal U}}
\def\cV{{\cal V}} \def\cW{{\cal W}} \def\cX{{\cal X}}
\def\cY{{\cal Y}} \def\cZ{{\cal Z}}
\def\bo {\bar{\o}}

\def\ua{\underline{\alpha}} \def\ubb{\underline{\beta}}
\def\ug{\underline{\gamma}}
\def\ub{\underline{\phantom{\alpha}}\!\!\!\beta}
\def\uc{\underline{\phantom{\alpha}}\!\!\!\gamma}
\def\um{\underline{\mu}} \def\un{\underline{\nu}}
\def\ud{\underline\delta}
\def\ue{\underline\epsilon}
\def\una{\underline a}\def\unA{\underline A}
\def\unb{\underline b}\def\unB{\underline B}
\def\unc{\underline c}\def\unC{\underline C}
\def\und{\underline d}\def\unD{\underline D}
\def\une{\underline e}\def\unE{\underline E}
\def\unf{\underline{\phantom{e}}\!\!\!\! f}\def\unF{\underline F}
\def\unm{\underline m}\def\unM{\underline M}
\def\unn{\underline n}\def\unN{\underline N}
\def\unp{\underline{\phantom{a}}\!\!\! p}\def\unP{\underline P}
\def\unq{\underline{\phantom{a}}\!\!\! q}
\def\unQ{\underline{\phantom{A}}\!\!\!\! Q}
\def\unH{\underline{H}}
\def\ul{\underline}

\def\As {{A \hspace{-6.4pt} \slash}\;}
\def\bs {{b \hspace{-6.4pt} \slash}\;}
\def\Ds {{D \hspace{-6.4pt} \slash}\;}
\def\ds {{\del \hspace{-6.4pt} \slash}\;}
\def\ss {{\s \hspace{-6.4pt} \slash}\;}
\def\ks {{ k \hspace{-6.4pt} \slash}\;}
\def\ps {{p \hspace{-6.4pt} \slash}\;}
\def\pas {{{p_1} \hspace{-6.4pt} \slash}\;}
\def\pbs {{{p_2} \hspace{-6.4pt} \slash}\;}

\def\Fh{\hat{F}}
\def\Vh{\hat{V}}
\def\Xh{\hat{X}}
\def\ah{\hat{a}}
\def\xh{\hat{x}}
\def\yh{\hat{y}}
\def\ph{\hat{p}}
\def\xih{\hat{\xi}}

\def\psit{\tilde{\psi}}
\def\Psit{\tilde{\Psi}}
\def\tht{\tilde{\th}}

\def\At{\tilde{A}}
\def\Qt{\tilde{Q}}
\def\Rt{\tilde{R}}
\def\Nt{\tilde{N}}

\def\at{\tilde{a}}
\def\st{\tilde{s}}
\def\ft{\tilde{f}}
\def\pt{\tilde{p}}
\def\qt{\tilde{q}}
\def\vt{\tilde{v}}
\def\nt{\tilde{n}}

\def\delb{\bar{\partial}}
\def\bz{\bar{z}}
\def\bD{\bar{D}}
\def\bB{\bar{B}}
\def\bo {\bar{\o}}

\def\bk{{\bf k}}
\def\bl{{\bf l}}
\def\bp{{\bf p}}
\def\bq{{\bf q}}
\def\br{{\bf r}}
\def\bx{{\bf x}}
\def\by{{\bf y}}
\def\bR{{\bf R}}
\def\bV{{\bf V}}
\def\bd{\begin{document}}

\def\ed{\end{document}}

\def\d{\delta}\def\D{\Delta}\def\ddt{\dot\delta}

\def\p{\partial} \def\del{\partial}
\def\xx{\times}
\def\uno{\mbox{1 \kern-.59em {\rm l}}}

\def\trp{^{\top}}
\def\inv{^{-1}}
\def\dag{{^{\dagger}}}
\def\pr{\prime}

\def\rar{\rightarrow}
\def\lar{\leftarrow}
\def\lrar{\leftrightarrow}

\def\cw{{\cal W}}
\def\cz{{\cal Z}}
\def\tcm{\tilde{\cal M}}
\def\sgn{{\rm sgn}}
\def\sd {d^{4|4}}
\def\lan{\langle}
\def\ran{\rangle}

\title{Hidden Conformal Symmetry and Quasinormal Modes}
\author{Bin Chen\\
Department of Physics,\\
and State Key Laboratory of Nuclear Physics and Technology,\\
and Center for High Energy Physics,\\
Peking University,\\
Beijing 100871, P.R. China\\
\email{bchen01@pku.edu.cn}}

\author{Jiang Long\\
Department of Physics,\\
Peking University,\\
Beijing 100871, P.R. China\\
\email{longjiang0301@gmail.com}}

\date{\today}

\abstract{We provide an algebraic way to calculate the quasinormal
modes of a black hole, which possesses  a hidden conformal symmetry.
We construct an infinite tower of quasinormal modes from the
highest-weight mode, in a simple and elegant way. For the scalar,
the hidden conformal symmetry manifests itself in the fact that the
scalar Laplacian could be rewritten in terms of the $SL(2,R)$
quadratic Casimir. For the vector and the tensor,  the hidden
conformal symmetry acts on them through Lie derivatives. We show
that for three-dimensional black holes, with an appropriate combination
of the components the radial equations of the vector and the tensor
could be written in
  terms of the Lie-induced quadratic Casimir. This makes the algebraic construction of the quasinormal modes feasible. Our results are in good agreement with the previous study.  }

\newpage

\begin{document}

\section{Introduction}

Very recently, motivated by the work in \cite{Andy2010}, it  has been
found that in many black holes which have holographic 2D CFT
descriptions there exists a hidden conformal symmetry. The hidden
conformal symmetry is realized by two sets of locally defined vector
fields $\{V_i, \bar{V}_i\}$ satisfying $SL(2,R)$ Lie algebra. This
symmetry is not globally defined, and is broken by the periodic
identification on angular variable. It could not be used to generate
new solutions. Nevertheless, it determines the scattering amplitudes
by acting on the solution space. More explicitly, the scalar
Laplacian could be written as the $SL(2,R)$ quadratic Casimir in
some limit region. The hidden conformal symmetry  was considered to
be essential to implement a holographic description of a black hole.
It was widely studied in various kinds of black holes, including the
4D Kerr-Newman\cite{Chen:2010xu}, 4D
Kerr-Newman-AdS-dS\cite{Chen:2010bh}, 3D black
holes\cite{Fareghbal:2010yd,{Li:2010zr}}, extremal black
holes\cite{Chen:2010fr} and others\cite{71}.

In retrospect, the appearance of hidden conformal symmetry is not a
surprise, considering the fact that the black hole is dual to a 2D
CFT. On the CFT side, the conformal symmetry restricts the form of
the correlation functions of the operators. Correspondingly, the
conformal symmetry acts on the solution space and determines the
scattering amplitudes. In the 4D Kerr case, the hidden conformal
symmetry only become manifest in the low frequency limit and in the
near region, but it determines not only the low frequency scattering
amplitudes but also  the super-radiant scattering ones. In 3D cases,
the hidden conformal symmetry generically manifests itself more
clearly, in all regions. Such features in 3D black holes are
due to the fact that the black holes are always locally isomorphic
to their covering spaces.

On the other hand, from AdS/CFT
correspondence\cite{Horowitz:1999jd,{Son:2002sd}} the quasinormal
modes, which determine the relaxation time of the perturbations
about the black hole, are related to the poles of the retarded
correlation function in the momentum space in the dual conformal
field theory. In a 2D CFT, the retarded Green's functions for an
operator with fixed conformal weights, fixed charges with respect to
chemical potentials are determined by the conformal
symmetry\cite{Cardy:1984rp}. The poles in the retarded Green's
function could be read easily. On the gravity side, the quasinormal
modes could be read from the eigenfunctions satisfying the purely
ingoing boundary condition at the black hole horizon and appropriate
boundary condition at the asymptotical
infinity\cite{quasinomral1999}. One has to solve the equations of
motion explicitly in order to get the eigenfunctions, whose analytic
forms are often out of reach. It is thus interesting to see that the
equation of motion with the hidden conformal symmetry acting on
could always be solved in terms of hypergeometric functions, due to
mathematical fact that the hypergeometric functions could form the
representation of the $SL(2,R)$ group. As a result, the quasinormal
modes could be read exactly. Actually one aim of this paper is to
show that even without solving the equations of motion explicitly,
we can determine the quasinormal modes in an elegant algebraic way.

Another issue on hidden conformal symmetry is how it acts on the
vector and tensor fields. In all the studies in the literature, the
hidden conformal symmetry has kept being discussed in the scalar
equation of motion. As it is an intrinsic property of the black
hole, it should also act on the other kinds of perturbations. In
this paper, we address this issue. For the locally defined vector
fields, they act on the vector and the tensor fields via
Lie-derivatives. We show that for 3D black holes, the hidden
conformal symmetry acts on the vector and tensor fields in a subtle
way. We find that only after an appropriate combination, the
equations of motion of the vector and tensor fields could be written
as the  Lie-induced quadratic Casimir:
  \be
(\mathcal{L}^2+m_t^2)T_+=0 \ee
 where
 \be
\mathcal{L}^2\equiv -
\mathcal{L}_{V_0}\mathcal{L}_{V_0}+\frac{1}{2}(\mathcal{L}_{V_1}\mathcal{L}_{V_{-1}}+\mathcal{L}_{V_{-1}}\mathcal{L}_{V_1})
\ee is the Casimir commuting with the Lie-derivatives
$\mathcal{L}_{V_i}$, and $T_+$ is an appropriate superposition of
tensor components. Actually the scalar equation of motion could also
be cast into the same form (1.2).

The fact that the equations of motion of all perturbations could be
written as  (1.2) allows us to construct the quasinormal modes in
an uniform way. We start from the highest-weight mode, which not
only satisfies the equations of motion but also obeys the condition
\be \mathcal{L}_{V_1}\Psi^{(0)}=0 ,\hs{5ex}
\mathcal{L}_{V_0}\Psi^{(0)}=h_R\Psi^{(0)}, \ee then construct the
descendent modes \be \Psi^n=(\mathcal{L}_{V_{-1}})^n\Psi^{(0)}. \ee
 It is nice to find that the descendent modes
constitute an infinite tower of quasinormal modes.
We will show that all the information of quasinormal modes is
encoded in the hidden conformal symmetry. The frequencies of the
quasinormal modes take the following form:   \be
\lambda_1\omega_R^{(n)}=\lambda_2 k+i(h_R+n_R), \hs{3ex}
\bar{\lambda}_1\omega_L^{(n)}=\bar{\lambda}_2 k+i(h_L+n_L)\ee where
$n_{L,R}$ are non-negative integers and $\l_i,\bar{\l}_i$ are
parameters in the hidden conformal symmetry. The spectrum of all
kinds of quasinormal modes share the same structure, with the
difference being from the conformal weights which are decided by the
$m^2_t$ term.

The way we approach the quasinormal modes is partly motivated by
the work in \cite{Sachs08}. In this paper, Ivo Sachs and Sergey N.
Solodukhin showed that quasinormal modes of the BTZ black hole in
topologically massive gravity may be derived from the Killing vector
fields. The essential aspect is that the Killing vectors form a
 $SL(2,R)$ Lie algebra locally so that they can build an
infinite tower of quasinormal modes. Our treatment is in spirit
similar to theirs, but differs in detail. In particular, our
investigation on the vector and tensor fields has not been presented
anywhere else before, to our knowledge. Moreover our discussion
includes the warped AdS$_3$ black hole and self-dual warped AdS$_3$
black hole of topological massive gravity, whose hidden conformal
symmetry is nontrivial, in contrast to  the BTZ black hole, which is
locally isomorphic to AdS$_3$ so that the hidden conformal symmetry
is not a real surprise. Actually the equations of motion in the
warped spacetime are of the form \be
(\mathcal{L}^2+b\bar{\mathcal{L}}_{\bar{V_0}}^2+m_t^2)T_+=0, \ee
which is slightly different from (1.2) but still allows us to
construct the quasinormal modes in the similar way. But now the
conformal weight depends not only on the mass but also on the extra
quantum numbers.

In the next section, we briefly review the realization of hidden
conformal symmetry. In Sec. III, we study the scalar perturbation
and determine the quasinormal modes as a warm-up. In Sec. IV, we
investigate the action of the hidden conformal symmetry on  the
vector and gravitational perturbations. In Sec. V, we discuss the
quasinormal modes of the BTZ black hole, and reproduce the well-known
results. In Sec. VI, we try to generalize the method to the warped
$AdS_3$ and self-dual warped $AdS_3$ black holes, which need a minor
modification of our construction. We will end with discussions in
Sec. VII. Some technical details are put into two appendixes.

\section{Hidden Conformal Symmetry}

In this paper,we will restrict to  generic nonextremal black holes
which have the hidden conformal symmetry. Generically we may
introduce the vector fields \bea
V_0&=&\lambda_1\partial_t+\lambda_2\partial_\phi,\nn\\
V_1&=&e^{\mu_1 t+\mu_2 \phi}[(A \frac{\Delta'}{\sqrt{\Delta}}+B \frac{1}{\sqrt{\Delta}}) \partial_t+(C \frac{\Delta'}{\sqrt{\Delta}}+D \frac{1}{\sqrt{\Delta}})\partial_\phi+\sqrt{\Delta}\partial_r],\label{conformal1}\\
V_{-1}&=&e^{-\mu_1 t-\mu_2 \phi}[(A\frac{\Delta'}{\sqrt{\Delta}}+B\frac{1}{\sqrt{\Delta}}) \partial_t+(C \frac{\Delta'}{\sqrt{\Delta}}+D \frac{1}{\sqrt{\Delta}})\partial_\phi-\sqrt{\Delta} \partial_r],\nn
\eea
where $\lambda_1,\lambda_2,\mu_1,\mu_2,A,B,C,D$ are all constants satisfying
\bea
\lambda_1\mu_1+\lambda_2\mu_2=-1,\nn\\
\lambda_1=2A,\nn\\
\lambda_2=2C,\\
\mu_1 B+\mu_2 D=0,\nn \eea and $\Delta=(r-r_+)(r-r_-)$,
$\Delta'=\frac{d\Delta}{dr}$. The above vector fields form a
$SL(2,R)$ algebra. \be [V_0,V_{\pm 1}]=\mp V_{\pm 1}, \hs{3ex}
[V_{+1},V_{-1}]=2V_0 \ee And similarly we can define the left sector
$\bar{V}_0,\bar{V}_{\pm1}$ with parameters $\bar{\mu}_i,
\bar{\l}_i,\bar{A},\bar{B},\bar{C},\bar{D}$.

The essential aspect is that the scalar Laplacian can be written as
the $SL(2,R)$ quadratic Casimir. More explicitly, the radial scalar
field equation in a black hole with holographic description is of
the form \be (V^2+m_s^2)\Phi(r)=0, \label{vscalar} \ee where
$V^2=-V_0^2+\frac{1}{2}(V_1V_{-1}+V_{-1}V_1)$ is the $SL(2,R)$
quadratic Casimir operator and $m_s^2$ is a constant. This is true
for the 4D Kerr(-Newman) black hole  in the low frequency and the near
region, and is always true for 3D black holes in the whole region.
Actually, one can give the explicit form of the Casimir. But we
would not give it here, instead we will derive it in the next
section in the general framework of Lie-derivative operation.

\section{Scalar Modes}

In this section we will derive the scalar equation using the Lie
derivatives. This seems useless since we have known the results in
Sec. II. But we will see that it is valuable to reproduce it in
another way, which could be generalized to discuss the vector and
tensor modes.

First we define Lie-induced quadratic Casimir \be
\mathcal{L}^2\equiv -
\mathcal{L}_{V_0}\mathcal{L}_{V_0}+\frac{1}{2}(\mathcal{L}_{V_1}\mathcal{L}_{V_{-1}}+\mathcal{L}_{V_{-1}}\mathcal{L}_{V_1})\label{lieCasimir}
\ee where $\mathcal{L}_{V_i},i=0,\pm1$ are the Lie derivatives with
respect to the vector fields $V_i$. $\mathcal{L}^2$ is analogue to
the $SL(2,R)$ quadratic Casimir $V^2$.

Let $\Phi$ be a scalar field and we immediately have
\be
\mathcal{L}^2\Phi=
\Pi^{\rho\sigma}\partial_\rho\partial_\sigma\Phi+\Sigma^{\rho}\partial_\rho\Phi
\ee
where we have defined
\bea
\Pi^{\rho\sigma}&\equiv &\frac{1}{2}(V_1^\rho V_{-1}^\sigma+V_1^\sigma V_{-1}^\rho)-V_0^\rho V_0^\sigma,  \label{Pi}\\
\Sigma^\rho&\equiv &\frac{1}{2}(V_1^\sigma\partial_\sigma
V_{-1}^\rho+V_{-1}^\sigma\partial_\sigma
V_1^\rho)-V_0^\sigma\partial_\sigma V_0^\rho.\label{Sigma} \eea The
explicit expressions of $\Pi$'s and $\Sigma$'s can be found in
Appendix A. We use them to find \be
\mathcal{L}^2\Phi=-\partial_r\Delta\partial_r
\Phi+[\frac{1}{(r-r_+)(r_+-r_-)}\sigma_+^2-\frac{1}{(r-r_-)(r_+-r_-)}\sigma_-^2]\Phi
\ee where $\sigma_\pm=(\pm A(r_+-r_-)+B)\partial_t+(\pm
C(r_+-r_-)+D)\partial_\phi$. Since we focus on the black holes which
have a hidden conformal symmetry, the scalar equation can be written
formally as \be (\mathcal{L}^2+m_s^2)\Phi=0 \label{lscalar} \ee
where $m_s$ is a constant which is related to the conformal weight
of the scalar. It varies for different black holes. Certainly for
the scalar, (\ref{lscalar}) is exactly the same as (\ref{vscalar}).

 To construct the tower of scalar quasinormal modes, we first impose the  condition:
\be \mathcal{L}_{V_1}\Phi^{(0)}=0 ,\hs{5ex}
\mathcal{L}_{V_0}\Phi^{(0)}=h_R\Phi^{(0)}\label{highestweight} \ee
to define the ``highest-weight'' mode.  Since \be
[\mathcal{L}_X,\mathcal{L}_Y]=\mathcal{L}_{[X,Y]},\hs{3ex}\mathcal{L}_{aX}=a\mathcal{L}_X
\label{lie} \ee where $X,Y$ are arbitrary vectors and $a$ is an
arbitrary constant, we get the following relation from the scalar
Eq. (\ref{lscalar}): \be h_R^2-h_R-m_s^2=0. \ee This determines
the conformal weight \be h_R=\frac{1}{2}(1+\sqrt{1+4m_s^2}). \ee We
have chosen the ``$+$" root to simplify our discussion. But the other
choice can also be considered easily.

From the mode $\Phi^{(0)}$, we construct an infinite tower of
quasinormal scalar modes $\Phi^{(n)}$ as \be
\Phi^{(n)}=(\mathcal{L}_{V_{-1}})^n\Phi^{(0)}, \hs{3ex}n\in N. \ee
All the $\Phi^{(n)}$ are descendents of the mode $\Phi^{(0)}$. Since
the Casimir $\mathcal{L}^2$ commutes with $\mathcal{L}_{V_i}$, $
i=0,\pm1$, $\Phi^{(n)}$ satisfy the scalar equation as well. To
compute the frequency of the quasinormal modes, we may expand the
scalar as \be \Phi=e^{-i\o t+i k \phi}\varphi, \ee as $\p_t$ and
$\p_\phi$ are always the Killing vectors of the black holes. For the
highest-weight mode $\Phi^{(0)}$, we have
\begin{equation}
\lambda_1\omega_0-\lambda_2 k_0=ih_R,
\end{equation}
where $\omega_0$ and $k_0$ are its frequency and  angular momentum.
In principle,  $k_0$ could be complex in the solution. Taking the
highest mode as quasinormal modes require $k_0$ be real. For the
descendent mode $\Phi^{(n)}$, we have \be
\mathcal{L}_{V_0}\Phi^{(n)}=(-i\lambda_1\omega_R^{(n)}+i\lambda_2
k^{(n)})\Phi^{(n)}, \ee where its frequency $\omega_R^{(n)}$ and angular
momentum $k_R^{(n)}$ are related to $\o_0$ and $k_0$ via the
relation
\begin{equation}
\omega_R^{(n)}=\omega_0-in\mu_1,\hs{5ex}
k_R^{(n)}=k_0+in\mu_2.\label{relationRk}
\end{equation}
To be a well-defined quasinormal mode, the angular momentum
$k_R^{(n)}$ should be real, which requires a choice of complex
$k_0$. Note that the real part of the $k_0$ and $k_R^{(n)}$ are
always the same, taken as $k$. From the relation (\ref{relationRk})
and the first relation in (2.2),  we obtain    \be
\lambda_1\omega_R^{(n)}=\lambda_2 k+i(h_R+n). \label{qnscalar} \ee
Alternatively it is more convenient to use just the algebraic
relation (\ref{lie})  to get this relation.

 The relation (\ref{qnscalar}) gives the frequencies of the scalar
quasinormal modes. We find that the frequencies of the modes only
depend on the parameters which appear in the hidden conformal
symmetry. Our construction relates the hidden conformal symmetry to
the structure of quasinormal modes directly.

Note that we can also determine the left sector modes from the other set of vector fields $\{\bar{V}_i\}$
according to the following rules:\\
(i) $R\to L$ \\
(ii)$\lambda_i\to \bar{\lambda}_i,{\mu}_i\to \bar{\mu}_i$, where
$i=1,2$.

In the next section, we will see that the similar construction could
be applied to the vector and gravitational modes, with subtle
modifications.

One can solve the highest-weight condition (\ref{highestweight})
explicitly. The solution is just \be
\Phi^{(0)}=C_0(r-r_+)^{-a-\frac{b}{r_+-r_-}}(r-r_-)^{-a+\frac{b}{r_+-r_-}},
\ee where $C_0$ is a integration constant and \bea
a&=&-iA\o+iCk, \nn\\
b&=&-iB\o+iD k. \nn \eea To satisfy the ingoing boundary condition
at the horizon $r=r_+$, we need \be A+\frac{B}{r_++r_-} <0. \ee We
will see that this is indeed the case for the black holes studied in
this paper. Asymptotically, the solution behaves as \be \Phi^{(0)}
\sim r^{-h_R}. \ee So we see that the solution has the right
behavior as the quasinormal mode. It is easy to find that the other
quasinormal modes have the same asymptotical behavior.

\section{Vector and Tensor Modes}

Let us first consider  the vector modes. Motivated by the
impressive result on scalar modes, we try to compute
$\mathcal{L}^2A_\mu$ and expect a similar structure. However it
turns out to be more complicated: \be
\mathcal{L}^2A_\mu=\Pi^{\rho\sigma}\partial_\rho\partial_\sigma
A_\mu+\Sigma^{\rho}\partial_\rho A_\mu+\partial_\mu \Sigma^{\sigma}
A_\sigma+\Upsilon_\mu^{\rho\sigma}\partial_\rho A_\sigma,
\label{lvector} \ee where $A_\mu$ is a vector field and
$\Pi^{\rho\sigma},\Sigma^\sigma$ are defined in (\ref{Pi}) and
(\ref{Sigma}). $\Upsilon_\mu^{\rho\sigma}$ is defined as \be
\Upsilon_\mu^{\rho\sigma}\equiv V_1^\rho
\partial_\mu V_{-1}^{\sigma}+V_{-1}^\rho\partial_\mu
V_1^\sigma-2V_0^\rho\partial_\mu V_0^\sigma. \ee At first glance,
(\ref{lvector}) looks quite different from  the scalar Eq.
(\ref{lscalar}). Especially the fact that the different components
are mixed together make things untractable. Nevertheless, we will
show that for 3D black holes, the relation (\ref{lvector}) could be
simplified. The detailed discussion on the vector and the tensor
perturbations in 3D black holes could be found in Appendix B.

Notice that the first and the second terms on the right-hand side of
(\ref{lvector}) are similar to the terms that appeared in the scalar
modes. The third term vanishes if we only consider the $A_t$ and
$A_\phi$ components since $\Sigma^\sigma$ is only a function of $r$
and independent of $t$ and $\phi$. To focus only on the $A_t$ and
$A_\phi$ components is plausible since the $A_r$ component can be
determined by the other components in 3D. The only trouble comes
from the fourth term, which cannot vanish automatically. The trick
is that we should consider the superposition of $A_t$ and $A_\phi$.
Let us define: \be A_+=\kappa_1A_t+\kappa_2A_\phi, \ee where
$\k_{1,2}$ are the constants to be determined. We find that a
suitable choice of $\kappa_1$ and $\kappa_2$ can make all the
components of $\Upsilon_\mu^{\rho\sigma}$ vanish. Actually, if \be
(\kappa_1\partial_t+\kappa_2\partial_\phi)V_i^{\sigma}=0 \ee where
$i=0,\pm1$ and $\sigma=t,\phi,r$, then
$\Upsilon_\mu^{\rho\sigma}=0$. The above condition can be satisfied
if \be \kappa_1:\kappa_2=-\mu_2:\mu_1. \ee Thus, we get \be
\mathcal{L}^2A_+=\Pi^{\rho\sigma}\partial_\rho\partial_\sigma
A_++\Sigma^{\rho}\partial_\rho A_+. \ee This shows that $A_+$
transform like a scalar. Now the question is if the equation of
$A_+$ could be written like a scalar: \be
(\mathcal{L}^2+m_v^2)A_+=0. \label{lvector2} \ee Certainly $m_v^2$
may be different from the scalar case, depending on the backgrounds
as well. We will show for the 3D black holes in this paper,
(\ref{lvector2}) is always true.

Next we turn to the tensor fields. For the tensor field $T_{\mu\nu}$, we have
\bea
\mathcal{L}^2 T_{\mu\nu}=\Pi^{\rho\sigma}\partial_\rho\partial_\sigma T_{\mu\nu}+\Sigma^{\rho}\partial_\rho T_{\mu\nu}+\partial_\mu \Sigma^\sigma T_{\sigma\nu}+\partial_\nu \Sigma^\sigma T_{\mu\sigma}
\nn\\
+\Xi_{\mu\nu}^{\rho\sigma}T_{\rho\sigma}+\Upsilon_\mu^{\rho\sigma}\partial_\rho
T_{\sigma\nu}+\Upsilon_\nu^{\rho\sigma}\partial_\rho
T_{\mu\sigma}\nn \eea where we have defined \be
\Xi_{\mu\nu}^{\rho\sigma}\equiv\partial_\mu V_1^\rho \partial_\nu
V_{-1}^\sigma+\partial_\nu V_1^\sigma \partial_\mu
V_{-1}^\rho-2\partial_\mu V_0^\rho \partial_\nu V_0^\sigma. \ee By
introducing \be
T_{+}=\kappa_1T_{tt}+\kappa_2T_{t\phi}+\kappa_3T_{\phi
t}+\kappa_4T_{\phi\phi}, \ee we find that when \bea
(\kappa_1\partial_t+\kappa_2\partial_\phi)V_i^{\sigma}=0,\nn\\
(\kappa_1\partial_t+\kappa_3\partial_\phi)V_i^{\sigma}=0,\nn\\
(\kappa_2\partial_t+\kappa_4\partial_\phi)V_i^{\sigma}=0,\label{tensor}\\
(\kappa_3\partial_t+\kappa_4\partial_\phi)V_i^{\sigma}=0,\nn \eea
all the redundant terms vanish and \be
\mathcal{L}^2T_{+}=\Pi^{\rho\sigma}\partial_\rho\partial_\sigma
T_{+}+\Sigma^{\rho}\partial_\rho T_{+}. \ee The condition
(\ref{tensor}) can be obeyed if the parameters $\mu_i,\k_i$ satisfy
the relations \be
\kappa_1:\kappa_2=-\mu_2:\mu_1=\kappa_3:\kappa_4,\hs{3ex}\kappa_2=\kappa_3.\label{kappa}
\ee As the vector case, we expect that the equations of motion of the 
tensor is \be (\mathcal{L}^2+m_t^2)T_{+}=0, \label{ltensor} \ee for
some constant $m_t$. We will show that for 3D black holes this is
the case in the next section.

The above construction may be generalized to the higher-rank tensor
fields. In general, for a rank $n$ tensor, we have \be
\mathcal{L}_{V}T_{l_1l_2\cdots
l_n}=V^{\mu}\partial_{\mu}T_{l_1l_2\cdots
l_n}+\partial_{l_1}V^{\lambda}T_{\lambda l_2\cdots
l_n}+\cdots+\partial_{l_n}V^{\lambda}T_{l_1 l_2\cdots
l_{n-1}\lambda}. \ee We can define a tensor as \be
T_+=\sum{\kappa_{\sigma_1\cdots\sigma_n}T_{\sigma_1\cdots\sigma_n}},
\ee where the summation is over all $\sigma_i=t,\phi$. Then we can
choose the $2^n$ coefficients $\kappa_{\cdots}$ such that \be
\mathcal{L}_{V_i}T_+=V_i^{\mu}\partial_{\mu}T_+ \ee with $i=0,\pm1$.
Note that this means that $T_+$ transform as a scalar under
$SL(2,R)$. This could be satisfied if \be
(\kappa_{\sigma_1\cdots\sigma_{j}t\sigma_{j+2}\cdots\sigma_{n}}\partial_t+\kappa_{\sigma_1\cdots\sigma_{j}\phi\sigma_{j+2}\cdots\sigma_{n}}\partial_{\phi})V^{\lambda}_i=0.
\ee There are $n\cdot 2^{n-1}$ constraints while there are
only $2^n$ degrees of freedom. But the above equations are not
independent and we can still determine the $2^n$ coefficients. One
can begin with $\kappa_{tt\cdots t}$ and end with
$\kappa_{\phi\phi\cdots\phi}$ step by step. Then one finds that \be
\mathcal{L}^2T_+=(\Pi^{\rho\sigma}\partial_\rho\partial_\sigma+\Sigma^{\rho}\partial_\rho)
T_+ \ee and we wish that the equation of motion of $T_+$ could be
written as \be (\mathcal{L}^2+m_{hs}^2)T_+=0 \ee with $m_{hs}$ being
a constant. In this paper, we just focus on the vector and rank $2$
tensor and leave the general case for a future study.

Before we go into the concrete examples, we would like to discuss
the physical implications of (\ref{lvector2}) and (\ref{ltensor}) on
the quasinormal modes. It is not hard to see that if we have the
relations (\ref{lvector2}) and (\ref{ltensor}), all the treatment on
the scalar modes could be applied to the vector and tensor modes.
That is to say, we can define  the ``highest-weight'' modes
$\Psi^{(0)}$, where $\Psi^{(0)}$ can be either $A_+$ or $T_{+}$, as
\be \mathcal{L}_{V_1}\Psi^{(0)}=0 ,\hs{5ex}
\mathcal{L}_{V_0}\Psi^{(0)}=h_R\Psi^{(0)}. \ee Moreover, since
(\ref{lie}) holds for arbitrary tensor fields, we can determine the
conformal weight to be $h_R=\frac{1}{2}(1+\sqrt{1+4m_i^2})$ with
$m_i^2=m_v^2$ or $m_t^2$. Similarly we can construct a tower of
quasinormal modes $\Psi^{(n)}$ as \be
\Psi^n=(\mathcal{L}_{V_{-1}})^n\Psi^{(0)}. \ee The frequency of the
quasinormal vector and tensor modes share the same structure as the
scalar modes (\ref{qnscalar}), with the difference coming from the
conformal weights. Certainly we can construct the left sector modes
in a similar way.

\section{quasinormal Modes in BTZ Black Hole}

In this section, we take the BTZ black hole as a typical example to
illustrate the above constructions of quasinormal modes. The
scalar, vector and spinor quasinormal modes of the BTZ black hole were
discussed in \cite{Briminghan01,{Cardoso:2001hn}}, while the massive
gravitational one in TMG theory was studied in \cite{Sachs08}(see
also \cite{Afshar:2010ii}). The metric of a BTZ black hole
is\cite{BTZ} \be
ds^2=-\frac{(r^2-r_+^2)(r^2-r_-^2)}{r^2}dt^2+\frac{r^2}{(r^2-r_+^2)(r^2-r_-^2)}dr^2+r^2(d\phi-\frac{r_+r_-}{r^2}dt)^2
\ee The left and right moving temperature are \be
T_L=\frac{r_+-r_-}{2\pi}, \hs{3ex}T_R=\frac{r_++r_-}{2\pi} \ee

 From the scalar equation we find  the hidden conformal symmetry in the BTZ black hole.
 In the BTZ case, we should replacing $r$ to $r^2$ and $\partial_r$ to $\partial_{r^2}$ in
 the conformal coordinates and the vector fields defined in (\ref{conformal1}).
 It turns out the parameters in (\ref{conformal1}) should be
\bea
\lambda_1=-\lambda_2=-\frac{1}{4\pi T_R},\hs{3ex}\mu_1=-\mu_2=2\pi T_R,\nn\\
A=-C=-\frac{1}{8\pi T_R},\hs{3ex}B=D=-\frac{\pi T_R}{2}\nn\\
\bar{\lambda_1}=\bar{\lambda_2}=-\frac{1}{4\pi T_L},\hs{3ex}\bar{\mu_1}=\bar{\mu_2}=2\pi T_L,\\
\bar{A}=\bar{C}=-\frac{1}{8\pi T_L},\hs{3ex}\bar{B}=-\bar{D}=-\frac{\pi T_L}{2}\nn
\eea
and we can also find that
\be
m_s^2=\frac{1}{4}m^2
\ee
where $m$ is the scalar mass. By substituting $\lambda_i$ into (\ref{qnscalar}), we find
\be
\omega_R^{(n)}=-k-i4\pi T_R(n_R+h_R),\hs{3ex}\omega_L^{(n)}=k-i4\pi T_L(n_L+h_L),\hs{3ex}n_L,n_R\in\mathbb{N}\label{BTZscalar}
\ee
where $h_L=h_R=\frac{1}{2}(1+\sqrt{1+m^2})$. This is in complete agreement with \cite{Briminghan01}.

To check the vector modes, we begin with the vector field equation
in three-dimensional spacetime: \be
\epsilon_{\lambda}^{\a\b}\partial_{\alpha}A_{\beta}=-mA_{\lambda}.
\ee We can show that for the BTZ black hole, the above equation can
be written as \bea
\tilde{\Delta}A_t=m^2A_t+2mA_{\phi},\\
\tilde{\Delta}A_{\phi}=m^2A_{\phi}+2mA_t.\nn \eea where we have
defined the operator
$\tilde{\Delta}=\frac{1}{\sqrt{-g}}\partial_{\mu}\sqrt{-g}g^{\mu\nu}\partial_{\nu}$,
which is an analogue to the Laplacian operator acting on the scalar
field.  See Appendix B.1 for more details. We immediately get \be
\tilde{\Delta}A_{\pm}=(m^2\pm2m)A_\pm \ee where $A_\pm=A_t\pm
A_{\phi}$. Note that this is just what we want. Using the language
in the above section,  as $-\mu_2:\mu_1=1$ and
$-\bar{\mu}_2:\bar{\mu}_1=-1$, we may choose $\kappa_1=\kappa_2=1$
and $\bar{\kappa}_1=-\bar{\kappa}_2=1$. Hence $A_\pm$  transform
like a scalar mode and \be m_v^2=\frac{1}{4}(m^2+2m) \ee for the
right-moving sector and \be \bar{m}_v^2=\frac{1}{4}(m^2-2m) \ee for
the left-moving sector. Then we get \be h_R=\frac{m}{2}+1, \hs{3ex}
h_L=\frac{m}{2} \ee which is  in agreement with the general result
that $|h_L-h_R|=s$. The frequencies of the quasinormal vector modes
are still given by (\ref{BTZscalar}).

Next, we turn to the gravitational modes. For the standard 3D
gravity, there is no propagating gravitational mode. However, for
the topological massive gravity, there is a massive graviton, whose
equation of motion could be written as a linear
equation\cite{Sachs08} \be
\epsilon_{\mu}^{\alpha\beta}\nabla_{\alpha}h_{\beta\nu}+mh_{\mu\nu}=0.
\ee Analogue to the vector mode, we can show that for the BTZ black
hole, \bea
\tilde{\Delta}h_{tt}&=&m^2h_{tt}+2mh_{\phi t}+2mh_{t\phi}+h_{tt}+2h_{\phi\phi},\nn\\
\tilde{\Delta}h_{t\phi}&=&m^2h_{t\phi}+2mh_{\phi\phi}+2mh_{tt}+h_{t\phi}+2h_{\phi t},\\
\tilde{\Delta}h_{\phi\phi}&=&m^2h_{\phi\phi}+2mh_{t\phi}+2mh_{\phi
t}+h_{\phi\phi}+2h_{tt}.\nn \eea See Appendix B for more detail.
After defining $h_{\pm}=h_{tt}\pm h_{t\phi}\pm h_{\phi
t}+h_{\phi\phi}$, we get \be \tilde{\Delta}h_{\pm}=(m^2\pm
4m+3)h_{\pm} \ee The above equations are precisely what we expect.
Since in this case, we should choose
$\kappa_1=\kappa_2=\kappa_3=\kappa_4=1$ and
$\bar{\kappa}_1=-\bar{\kappa}_2=-\bar{\kappa}_3=\bar{\kappa}_4=1$.
Then $h_\pm$ are just the $T_{\pm}$ we have defined in the previous
section. Consequently, we find that \be m_t^2=\frac{1}{4}(m^2+4m+3)
\ee for the right-moving sector and \be
\bar{m}_t^2=\frac{1}{4}(m^2-4m+3) \ee for the left-moving sector.
The right and left conformal weight are respectively \be
h_R=\frac{m+3}{2},\hs{3ex}h_L=\frac{m-1}{2}, \label{tensorweight}\ee
which again is consistent with the fact that $|h_L-h_R|=s$. The
frequencies of the gravitational quasinormal modes take the same
form as (\ref{BTZscalar}).

\section{quasinormal Modes in Warped $AdS_3$ and Self-dual Warped $AdS_3$  Black Hole}

In this section, we will generalize the algebraic method to the
warped $AdS_3$ and the self-dual warped $Ad S_3$ black holes. For
the warped black holes, the scalar equations could not be simply
written as (\ref{lscalar}). Actually they take 
 the following form:
\be (\mathcal{L}^2+b\bar{\mathcal{L}}_{\bar{V}_0}^2+m_s^2)\Phi=0 \ee
with $b$ and $m_s^2$ being constants. This is a little different
from the previous discussion due to the presence of the 
$b\bar{\mathcal{L}}_{\bar{V_0}}^2$ term. Nevertheless, we can still
construct a tower of right-moving modes by imposing the conditions
\be
\mathcal{L}_{V_1}\Phi^{(0)}=0,\hs{3ex}\mathcal{L}_{V_0}\Phi^{(0)}=h_R\Phi^{(0)},\hs{3ex}\Phi^{(n)}=(\mathcal{L}_{V_{-1}})^n\Phi^{(0)}.
\ee The first two conditions  just define the ``highest-weight''
mode. And the last equation construct the descendent modes. Because of 
the commutative relation
$[\mathcal{L}_{V_{-1}},\bar{\mathcal{L}}_{\bar{V_0}}]=0$, all the
modes $\Phi^{(n)}$ satisfy the scalar equation as well. The
following discussion is similar to the one in Sec. III. Here we
only give the results: \be
h_R=\frac{1}{2}(1+\sqrt{1+4(bq^2+m_s^2)}),\hs{3ex}
\lambda_1\omega_R^{(n)}=\lambda_2 k+i(h_R+n) \ee where $q$ is
defined by $\bar{\mathcal{L}}_{\bar{V_0}}\Phi^{(0)}=q\Phi^{(0)}$.
However,we can not construct the left-moving modes due to the
noncommutative relation of $\bar{\mathcal{L}}_{\bar{V_0}}$ and
$\bar{\mathcal{L}}_{\bar{V}_{-1}}$. Because of the presence of
$b\bar{\mathcal{L}}_{\bar{V_0}}^2$ term in the scalar equation, the
conformal weight  depends on the quantum number $q$. This fact is in
consistency with the known result.

Next, we try to generalize the above discussion to the vector modes.
In this case, we find that for any vector $A_{\mu}$ \be
\bar{\mathcal{L}}_{\bar{V_0}}^2A_{\mu}=\bar{V}_0^{\rho}\partial_{\rho}\bar{V}_0^{\sigma}\partial_{\sigma}A_{\mu}.
\ee All the redundant term of $\partial_{\mu}\bar{V}_0^{\lambda}$
vanish since $\bar{V}_0^{\lambda}$ are constant numbers. This
implies that we can still define $A_+=\kappa_1A_t+\kappa_2A_{\phi}$
with $\kappa_1:\kappa_2=-\mu_2:\mu_1$. We still expect that it
transforms as a scalar. More explicitly, we wish \be
(\mathcal{L}^2+b\bar{\mathcal{L}}_{\bar{V_0}}^2+m_v^2)A_+=0. \ee If
this is true, we can discuss the vector modes parallel to the
treatment on the scalar modes. We will check this point in the
warped $AdS_3$ and self-dual warped $AdS_3$ black hole backgrounds
in the next two subsections.

For the warped spacetime, the equation of motion of the
gravitational mode could not be written as a linear
equation\cite{Anninos:2009zi} and is much more involved. Here we
just assume that there is a massive rank 2 symmetric tensor mode in
the warped spacetime. In 3D dimension, its equation of motion is \be
\epsilon_{\mu}^{\alpha\beta}\nabla_{\alpha}h_{\beta\nu}+mh_{\mu\nu}=0.
\ee In this case, we can still define
$T_+=\kappa_1T_{tt}+\kappa_2T_{t\phi}+\kappa_3T_{\phi
t}+\kappa_4T_{\phi\phi}$ with $\k_i$'s satisfying (\ref{kappa}) and
wish it to satisfy the equation of the form \be
(\mathcal{L}^2+b\bar{\mathcal{L}}_{\bar{V_0}}^2+m_t^2)T_+=0. \ee If
this is true, it allows us to construct the tensor quasinormal
modes in the similar way.

\subsection{ Warped $AdS_3$ black hole}

 The metric of the spacelike stretched warped $AdS_3$ black hole  is\cite{Andy08}
\be ds^2=dt^2+2M(r)dtd\phi+N(r)d\phi^2+Q(r)dr^2 \ee where \bea
M&=&\nu r-\frac{1}{2}\sqrt{r_+r_-(\nu^2+3)},\nn\\
N&=&\frac{r}{4}[3(\nu^2-1)r+(\nu^2+3)(r_++r_-)-4\nu\sqrt{r_+r_-(\nu^2+3)}],\label{warpedmetric}\\
Q&=&\frac{1}{(\nu^2+3)(r-r_+)(r-r_-)}.\nn \eea From warped AdS/CFT
correspondence,  the right- and left-moving temperatures in the dual
2D CFT are \be
T_L=\frac{\nu^2+3}{8\pi}(r_++r_--\frac{\sqrt{r_+r_-(\nu^2+3)}}{\nu}),\hs{3ex}T_R=\frac{(\nu^2+3)(r_+-r_-)}{8\pi}.
\ee The hidden conformal symmetry of the warped AdS$_3$ black hole has
been discussed in \cite{Fareghbal:2010yd}. From the scalar equation,
we find that \bea
\lambda_1=-\frac{2\nu T_L}{(\nu^2+3)T_R},\hs{3ex}\lambda_2=\frac{1}{2\pi T_R},\hs{3ex}\mu_1=0,\hs{3ex}\mu_2=-2\pi T_R\nn\\
A=-\frac{\nu T_L}{(\nu^2+3)T_R},\hs{3ex}B=-\frac{\nu(r_+-r_-)}{\nu^2+3},\hs{3ex}C=\frac{1}{4\pi T_R},\hs{3ex}D=0\nn\\
\bar{\lambda}_1=-\frac{2\nu}{\nu^2+3},\hs{3ex}\bar{\lambda}_2=0,\hs{3ex}\bar{\mu}_1=\frac{\nu^2+3}{2\nu},\hs{3ex}\bar{\mu}_2=2\pi T_L\\
\bar{A}=-\frac{\nu}{\nu^2+3},\hs{3ex}\bar{B}=\frac{\nu8\pi
T_L}{(\nu^2+3)^2},\hs{3ex}\bar{C}=0,\hs{3ex}\bar{D}=\frac{2}{\nu^2+3}\nn
\eea and $b,q,m_s^2$ are \be
b=\frac{3(\nu^2-1)}{4\nu^2},\hs{3ex}q=i\frac{2\nu\omega}{\nu^2+3},\hs{3ex}m_s^2=\frac{m^2}{\nu^2+3}
\ee where $m$ is the scalar mass. Hence the scalar conformal weight
is \be
h_R=\frac{1}{2}(1+\sqrt{1+4\frac{m^2(\nu^2+3)-3(\nu^2-1)\omega^2}{(\nu^2+3)^2}})
\ee As emphasized in \cite{ChenXu2}, to compare with the poles of
the correlation functions in the dual CFT, we should take the
following identification on quantum numbers into
account\cite{ChenXu2} \be
\tilde{\omega}=\frac{2k}{\nu^2+3},\hs{3ex}\tilde{k}=\frac{2\nu\omega}{\nu^2+3},
\ee where $\tilde{\o},\tilde{k}$ are the quantum numbers of global
warped AdS$_3$ spacetime. Then we find the scalar quasinormal modes
with the frequencies\be
\tilde{\omega}_R^{(n)}=\frac{1}{\nu^2+3}(-4\pi T_L\tilde{k}-i4\pi
T_R(n+h_R)). \label{QNwarped}\ee This is in agreement with the
result in \cite{{ChenXu09},ChenXu2}.

Next, we check the vector modes. Since in this case, $\mu_1=0$
indicates $\kappa_1=1, \kappa_2=0$, we should choose $A_+=A_t$. In
Appendix B.1 we show that $A_t$ satisfy \be
\tilde{\Delta}A_t=(m^2+2m\nu)A_t, \label{warpedvector} \ee which
could be rewritten   as \be
(\mathcal{L}^2+b\bar{\mathcal{L}}_{\bar{V_0}}^2+m_v^2)A_+=0 \ee with
$m_v^2=\frac{m^2+2m\nu}{\nu^2+3}$ and $b$ has been given in the scalar case. This is
in agreement with our expectation. Hence, the vector conformal
weight is \be
h_R=\frac{1}{2}(1+\sqrt{1+4(\frac{(m^2+2m\nu)}{\nu^2+3}-\frac{3(\nu^2-1)\tilde{k}^2}{4\nu^2})})
\ee where we have used the identification
$\tilde{k}=\frac{2\nu\omega}{\nu^2+3}$. The result is in perfect
match with the result in \cite{ChenXu2}. The spectrum of the vector
quasinormal modes takes the same form as (\ref{QNwarped}).

For the tensor mode, as $\mu_1=0, \mu_2\neq 0$, we may choose
 \be
\k_1=1, \k_2=\k_3=\k_4=0 \ee and have \be T_+=h_{tt}.\ee From the
equation of motion, we learn that \be
\tilde{\Delta}h_{tt}=(m^2+4m\nu+3\nu^2)h_{tt} \ee which could be
rewritten as \be
(\mathcal{L}^2+b\bar{\mathcal{L}}_{\bar{V_0}}^2+m_t^2)T_+=0 \ee with
$m_t^2=\frac{m^2+4m\nu+3\nu^2}{\nu^2+3}$. Thus we find the conformal weight of the
tensor mode \be
h_R=\frac{1}{2}(1+\sqrt{1+4(\frac{(m^2+4m\nu+3\nu^2)}{\nu^2+3}-\frac{3(\nu^2-1)\tilde{k}^2}{4\nu^2})}).
\ee When $\nu=1$, the warped black hole reduces to the BTZ black hole
and the tensor conformal weight reduces to $h_R$ in
(\ref{tensorweight}). The spectrum of tensor quasinormal modes takes
the same form as (\ref{QNwarped}).

\subsection{Self-dual warped $AdS_3$ black hole}

The self-dual warped AdS$_3$ black hole is a vacuum solution of 3D
topological massive gravity. It could be described by the metric
\bea \label{selfBH} ds^2 ~=~ \frac{l^2}{\nu^2 +
3}\left(-(r-r_+)(r-r_-)\,dt^2 + \frac{1}{(r-r_+)(r-r_-)}
\,dr^2\right.
\\\left.+\frac{4 \nu^2}{\nu^2 +3} ( \alpha d\phi + (r-\frac{r_+ + r_-}{2})\, dt)^2\right)\,,
\eea where the coordinates range as $t\in[-\infty,\infty]$,
${r}\in[-\infty,\infty]$ and $\phi \sim \phi + 2\pi$.

 The hidden conformal symmetry of this black hole has been discussed in \cite{Li:2010zr}\footnote{The
 scalar quasinormal mode of self-dual black hole has been discussed in \cite{Li:2010zr} in the similar way
 as the one in \cite{Sachs08}.}. It turns out that the parameters in the vector fields take the following values
\bea
\lambda_1=-\frac{1}{2\pi T_R},\hs{3ex}\lambda_2=0,\hs{3ex}\mu_1=2\pi T_R,\hs{3ex}\mu_2=0,\nn\\
A=-\frac{1}{4\pi T_R},\hs{3ex}B=0,\hs{3ex}C=0,\hs{3ex}D=\frac{T_R}{T_L},\nn\\
\bar{\lambda}_1=0,\hs{3ex}\bar{\lambda}_2=-\frac{1}{2\pi T_L},\hs{3ex}\bar{\mu}_1=0,\hs{3ex}\bar{\mu}_2=2\pi T_L,\\
\bar{A}=0,\hs{3ex}\bar{B}=1,\hs{3ex}\bar{C}=-\frac{1}{4\pi
T_L},\hs{3ex}\bar{D}=0,\nn \eea where the right- and left- moving
temperature are \be T_L=\frac{\alpha}{2\pi},
\hs{3ex}T_R=\frac{r_+-r_-}{4\pi}. \ee The $b,q,m_s^2$ are \be
b=\frac{3(\nu^2-1)}{4\nu^2},\hs{3ex}q=-\frac{ik}{\alpha},\hs{3ex}m_s^2=\frac{m^2}{\nu^2+3}. \ee The conformal weight for
the scalar is just \be
h_R=\frac{1}{2}(1+\sqrt{1-\frac{3(\nu^2-1)k^2}{\nu^2\alpha^2}+\frac{4m^2}{\nu^2+3}}).
\ee

For the vector field, as $\mu_2=0$, we may choose $\l_2=1$ so that
$A_+=A_{\phi}$. In Appendix B.1, we see that $A_{\phi}$ satisfy \be
\tilde{\Delta}A_{\phi}=(m^2-2m\nu)A_{\phi}, \ee which could be cast
into the form \be
(\mathcal{L}^2+b\bar{\mathcal{L}}_{\bar{V_0}}^2+m_v^2)A_+=0 \ee
 with $m^2_v=\frac{m^2-2m\nu}{\nu^2+3}$. The conformal weight for the vector is then
\be
h_R=\frac{1}{2}(1+\sqrt{1-\frac{3(\nu^2-1)k^2}{\nu^2\alpha^2}+\frac{4(m^2-2m\nu)}{\nu^2+3}}).
\ee

For the tensor field, as $\mu_2=0, \mu_1\neq 0$, we may choose
 \be
\k_4=1, \k_1=\k_2=\k_3=0 \ee and have \be T_+=h_{\phi\phi}.\ee From
the equation of motion, we learn that \be
\tilde{\Delta}h_{\phi\phi}=(m^2-4m\nu+3\nu^2)h_{tt} \ee which could
be rewritten as \be
(\mathcal{L}^2+b\bar{\mathcal{L}}_{\bar{V_0}}^2+m_t^2)T_+=0 \ee with
$m_t^2=\frac{m^2-4m\nu+3\nu^2}{\nu^2+3}$. The conformal weight of the tensor field
is just \be
h_R=\frac{1}{2}(1+\sqrt{1-\frac{3(\nu^2-1)k^2}{\nu^2\alpha^2}+\frac{4(m^2-4m\nu+3\nu^2)}{\nu^2+3}}),
\ee which could reduce to $h_L$ in (\ref{tensorweight}) at $\nu=1$.

In all cases, the quasinormal modes can be written as \be
\omega_R^{(n)}=-i2\pi T_R(n+h_R) \ee where $h_R$ can be  the scalar,
 the vector or the tensor conformal weight. The results all in good agreement
with \cite{Bin10selfdual,{Li:2010sv}}.

\section{Discussions}

In this paper, we have studied the relation between the hidden
conformal symmetry and the quasinormal modes. We found that the
spectrum of the quasinormal modes  may be directly read out from
the action of the hidden conformal symmetry on various
perturbations. Our construction provides a direct rule to find the
spectrum. The rule is simple and show clearly that the quasinormal
modes are determined completely by the hidden conformal symmetry. We
found that in the
 spectrums,
 \be
 \o \propto -i2\pi T(h+n),
 \ee
 which is in
accordance with the structure of the poles of the correlation
functions of the dual operators in CFT.

Our construction is based on the relation (\ref{lie}) on the
Lie-derivatives and the fact that the Lie-induced Casimir
$\mathcal{L}^2$ defined in (\ref{lieCasimir}) commutes with the
Lie-derivatives. Starting from the highest-weight mode, we can
construct its infinite tower of descendent modes. For the scalar,
the construction is straightforward, as shown in Sec. III. However,
the action of the hidden conformal symmetry on the vector and tensor
field is highly nontrivial. We observed that only after some
suitable composition the vector and the gravitation modes behaved
like the scalar modes. This allowed us to treat the scalar, vector
and gravitational modes in a uniform way. From our construction,
the spectrum of various kinds of quasinormal modes are in agreement
with the CFT prediction and previous study.  Moreover, our
discussion in Sec. IV suggested that our treatment could be
applied to higher-rank tensor fields. It would be nice to have a
detailed study on this question. Another interesting issue is to
study if the hidden conformal symmetry can determine the fermionic
quasinormal modes.

We applied our method to the case of the BTZ black hole and find perfect
agreement with the known results. For the warped AdS$_3$ and
self-dual warped AdS$_3$ black holes, the discussion is subtler.
Even for the scalar mode, the scalar equation could not be simply
written as the $SL(2,R)$ quadratic Casimir for all quantum numbers.
Nevertheless, we can still apply our treatment with a minor
modification. For all the scalar, vector and tensor modes, we
managed to construct towers of the quasinormal modes, in agreement
with the ones found in the literature. Strictly speaking, we only
succeeded in finding one set of the quasinormal modes,
corresponding to the poles of the correlation functions of the
right-moving sector in the dual CFT. It would be nice to find the
other set, corresponding to  the left-moving ones.

In this paper, we studied the quasinormal modes of the nonextremal
black holes. Since the coordinates that are used to implement the
hidden conformal symmetry are different in the extremal
case\cite{Chen:2010fr}, it is interesting to see if the same
construction works for the extreme black holes. We expect an 
similar conclusion.

We discussed the action of the hidden conformal symmetry on the
vector and tensor fields in three-dimensional spacetime. It would be
interesting to investigate this issue in the Kerr/CFT correspondence
in higher dimensions. However in this case, the problem is much more
complicated because we have to apply the Newman-Penrose
formalism to obtain the Teukolsky master
equation\cite{Hartman:2009nz,{Chen:2010ni}} of high spin
perturbations. It is not clear how the hidden conformal symmetry is
realized in this framework.

\section*{Acknowledgments}

The work was in part supported by NSFC Grant No. 10775002, 10975005.
We would like to thank for KITPC for hospitality, where this project
was initiated.

\section*{Appendix A}
The explicit forms of $\Pi^{\rho\sigma}$ and $\Sigma^\sigma$ are
 the following: \bea
\Pi^{rr}&=&-\Delta\nn\\
\Pi^{tt}&=&(A\frac{\Delta'}{\sqrt{\Delta}}+B\frac{1}{\sqrt{\Delta}})^2-\lambda_1^2\nn\\
\Pi^{\phi\phi}&=&(C\frac{\Delta'}{\sqrt{\Delta}}+D\frac{1}{\sqrt{\Delta}})^2-\lambda_2^2\\
\Pi^{t\phi}&=&\Pi^{\phi t}=(A\frac{\Delta'}{\sqrt{\Delta}}B\frac{1}{\sqrt{\Delta}})(C\frac{\Delta'}{\sqrt{\Delta}}+D\frac{1}{\sqrt{\Delta}})-\lambda_1\lambda_2\nn\\
\Pi^{rt}&=&\Pi^{rt}=\Pi^{r\phi}=\Pi^{\phi r}=0\nn \eea and \be
\Sigma^t=\Sigma^\phi=0,\hs{3ex} \Sigma^r=-\Delta'. \ee

\section*{Appendix B:Vector and Tensor Perturbation In (2+1)-dim. Black Holes}

In this section, we give a discussion of the vector and the tensor
perturbations  in (2+1)-dim. black holes. The discussion is not
restricted to the  black holes studied in this paper. In fact,we
only require the following conditions on the metric: \be
\partial_tg_{\mu\nu}=0,\hs{3ex}\partial_{\phi}g_{\mu\nu}=0,\hs{3ex}g_{rt}=g_{r\phi}=0.\label{metriccon}
\ee
\subsubsection*{B.1 Vector perturbation}

We begin with the vector equation in 3D spacetime \be
\epsilon_{\lambda}^{\a\b}\partial_{\alpha}A_{\beta}=-mA_{\lambda},
\ee which could be written in components  \bea
A_r&=&-\frac{1}{m}\epsilon_r^{t\phi}(\partial_rA_{\phi}-\partial_{\phi}A_t),\nn\\
\partial_rA_t&=&\partial_tA_r-m\frac{(\epsilon_{\phi}^{r\phi}A_t-\epsilon_t^{r\phi}A_{\phi})}{\epsilon_{\phi}^{r\phi}\epsilon_t^{rt}-\epsilon_t^{r\phi}\epsilon_{\phi}^{rt}},\\
\partial_rA_{\phi}&=&\partial_{\phi}A_r-m\frac{(\epsilon_{t}^{rt}A_{\phi}-\epsilon_{\phi}^{rt}A_{t})}{\epsilon_{\phi}^{r\phi}\epsilon_t^{rt}-\epsilon_t^{r\phi}\epsilon_{\phi}^{rt}}.\nn
\eea Obviously the $A_r$ component could be decided by $A_t$ and
$A_\phi$. Our goal is to find an equation which is analogue to the
scalar equation \be
\frac{1}{\sqrt{-g}}\partial_{\mu}\sqrt{-g}g^{\mu\nu}\partial_{\nu}\Phi=\cdots.
\ee This motivates us to compute
$\tilde{\Delta}A_i=\frac{1}{\sqrt{-g}}\partial_{\mu}\sqrt{-g}g^{\mu\nu}\partial_{\nu}A_i$,
with $i=t,\phi$. The results are \bea
\tilde{\Delta}A_t&=&m^2A_t+m\tilde{\epsilon}^{tr\phi}\frac{1}{\sqrt{-g}}(g_{\phi t}'A_t-g_{tt}'A_{\phi}),\nn\\
\tilde{\Delta}A_{\phi}&=&m^2A_{\phi}+m\tilde{\epsilon}^{\phi
rt}\frac{1}{\sqrt{-g}}(g_{t\phi}'A_{\phi}-g_{\phi\phi}'A_t),\nn \eea
where $\epsilon_{\lambda}^{\mu\nu}$ is the Levi-Civita tensor and
$\tilde{\epsilon}^{tr\phi}$ is the Levi-Civita symbol with
$\tilde{\epsilon}^{tr\phi}=1$.

For the BTZ black hole, all the $r$ coordinates should be replaced
by $r^2$, and then \be
g_{t\phi}'=0,\hs{3ex}g_{tt}'=-1,\hs{3ex}g_{\phi\phi}'=1,\hs{3ex}\sqrt{-g}=\frac{1}{2}.
\ee Note that the derivative should be taken with respect to $r^2$.
Then we get \bea
\tilde{\Delta}A_t=m^2A_t+2mA_{\phi},\\
\tilde{\Delta}A_{\phi}=m^2A_{\phi}+2mA_t.\nn \eea

For the spacelike stretched warped AdS$_3$ black hole, since
$g_{tt}'=0, g_{t\phi}'=\nu, \sqrt{-g}=\frac{1}{2}$, we find that \be
\tilde{\Delta}A_t=(m^2+2m\nu)A_t. \ee For the self-dual warped
AdS$_3$ black hole,
$g_{t\phi}'=\frac{4\nu^2\alpha}{(\nu^2+3)^2},g_{\phi\phi}'=0,\sqrt{-g}=\frac{2\nu \alpha}{(\nu^2+3)^2}$,
 then  $A_{\phi}$ satisfy \be
\tilde{\Delta}A_{\phi}=(m^2-2m\nu)A_{\phi}. \ee

\subsubsection*{B.2 Tensor perturbation}

In 3D spacetime, the rank 2 symmetric tensor perturbation obeys the
equation  \be
\epsilon_{\mu}^{\alpha\beta}\nabla_{\alpha}h_{\beta\nu}+mh_{\mu\nu}=0.
\ee For the BTZ black hole in 3D TMG theory, this is the equation
for a massive graviton. However for the warped AdS spacetime, the
gravitational perturbation could not be put into such a simple
form\cite{Anninos:2009zi}. Nevertheless we can still assume a
massive tensor perturbation in the backgrounds, satisfying this
equation.

From this equation,  we can easily find \bea
h_{\phi\phi}'&=&\partial_{\phi}h_{r\phi}+\Gamma_{(\phi\phi)}+m_{(\phi\phi)},\nn\\
h_{t\phi}'&=&\partial_t h_{r\phi}+\Gamma_{(t\phi)}+m_{(t\phi)},\nn\\
h_{\phi t}'&=&\partial_{\phi}h_{rt}+\Gamma_{(\phi t)}+m_{(\phi t)}, \nn\\
h_{tt}'&=&\partial_{t}h_{rt}+\Gamma_{(tt)}+m_{(tt)},\nn \eea
and \bea
h_{rt}&=&\frac{1}{W}(a_{rt}+m_{(rt)}),\nn\\
h_{r\phi}&=&\frac{1}{W}(a_{r\phi}+m_{(r\phi)}),\nn\\
h_{rr}&=&g_{rr}^2(\epsilon^{rt\phi})^2(g_{tt}h_{\phi\phi}-2g_{t\phi}h_{t\phi}+g_{\phi\phi}h_{tt}),\nn
\eea where we have defined \bea
\Gamma_{(\phi\phi)}&=&\Gamma^{\lambda}_{r\phi}h_{\phi\lambda}-\Gamma^{\lambda}_{\phi\phi}h_{r\lambda}\nn\\
\Gamma_{(t\phi)}&=&\Gamma^{\lambda}_{r\phi}h_{t\lambda}-\Gamma^{\lambda}_{t\phi}h_{r\lambda}\nn\\
\Gamma_{(\phi t)}&=&\Gamma^{\lambda}_{rt}h_{\phi\lambda}-\Gamma^{\lambda}_{\phi t}h_{r\lambda}\nn\\
\Gamma_{(tt)}&=&\Gamma^{\lambda}_{rt}h_{t\lambda}-\Gamma^{\lambda}_{tt}h_{r\lambda}\nn\\
m_{(\phi\phi)}&=&mg_{rr}\epsilon^{\phi rt}(g_{t\phi}h_{\phi\phi}-g_{\phi\phi}h_{t\phi})\nn\\
m_{(t\phi)}&=&mg_{rr}\epsilon^{tr\phi }(g_{\phi t}h_{\phi t}-g_{tt}h_{\phi\phi})\nn\\
m_{(\phi t)}&=&mg_{rr}\epsilon^{\phi rt}(g_{t\phi}h_{\phi t}-g_{\phi\phi}h_{tt})\\
m_{(tt)}&=&mg_{rr}\epsilon^{tr\phi }(g_{\phi t}h_{tt}-g_{tt}h_{\phi
t})\nn \\
m_{(rt)}&=&-mg_{rr}\epsilon^{rt\phi}(\partial_th_{\phi t}-\partial_{\phi}h_{tt})\nn\\
m_{(r\phi)}&=&-mg_{rr}\epsilon^{rt\phi}(\partial_th_{\phi
\phi}-\partial_{\phi}h_{t\phi})\nn\\
W&=&m^2-\frac{1}{4}(\epsilon^{rt\phi})^2(g_{t\phi}'g_{\phi
t}'-g_{tt}'g_{\phi\phi}')\nn
 \eea and \bea
a_{rt}&=&-\frac{1}{2}g_{rr}(\epsilon^{rt\phi})^2[-g_{tt}'(\partial_t h_{\phi\phi}-\partial_{\phi}h_{t\phi})+g_{t\phi}'(\partial_th_{\phi t}-\partial_{\phi}h_{tt})]\nn\\
a_{r\phi}&=&-\frac{1}{2}g_{rr}(\epsilon^{rt\phi})^2[-g_{\phi\phi}'(\partial_{\phi}
h_{tt}-\partial_{t}h_{\phi t})+g_{\phi
t}'(\partial_{\phi}h_{t\phi}-\partial_{t}h_{\phi\phi})].\nn \eea
Similar to the vector case, we find the following equations: \bea
\tilde{\Delta}h_{\phi\phi}&=&m^2h_{\phi\phi}+\frac{m\tilde{\epsilon}^{\phi rt}}{\sqrt{-g}}\alpha_{\phi\phi}+\frac{\beta}{4(-g)}h_{\phi\phi}+\frac{1}{2(-g)}\gamma_{\phi\phi}+(I)_{\phi\phi}+(II)_{\phi\phi},\nn\\
\tilde{\Delta}h_{t\phi}&=&m^2h_{t\phi}+\frac{m\tilde{\epsilon}^{\phi rt}}{\sqrt{-g}}\alpha_{t\phi}+\frac{\beta}{4(-g)}h_{t\phi}+\frac{1}{2(-g)}\gamma_{t\phi}+(I)_{t\phi}+(II)_{t\phi},\nn\\
\tilde{\Delta}h_{\phi t}&=&m^2h_{\phi t}+\frac{m\tilde{\epsilon}^{t r\phi}}{\sqrt{-g}}\alpha_{\phi t}+\frac{\beta}{4(-g)}h_{\phi t}+\frac{1}{2(-g)}\gamma_{\phi t}+(I)_{\phi t}+(II)_{\phi t},\nn\\
\tilde{\Delta}h_{tt}&=&m^2h_{tt}+\frac{m\tilde{\epsilon}^{t
r\phi}}{\sqrt{-g}}\alpha_{tt}+\frac{\beta}{4(-g)}h_{tt}+\frac{1}{2(-g)}\gamma_{tt}+(I)_{tt}+(II)_{tt},\nn
\eea where \bea
\alpha_{\phi\phi}&=&2g_{t\phi}'h_{\phi\phi}-2g_{\phi\phi}'h_{t\phi}\nn\\
\alpha_{t\phi}&=&-g_{\phi\phi}'h_{tt}+g_{tt}'h_{\phi\phi}\nn\\
\alpha_{\phi t}&=&-g_{tt}'h_{\phi\phi}+g_{\phi\phi}'h_{tt}\nn\\
\alpha_{\phi\phi}&=&2g_{\phi t}'h_{tt}-2g_{tt}'h_{\phi t}\nn \eea
\bea
\gamma_{\phi\phi}&=&g_{\phi t}'g_{t\phi}'h_{\phi\phi}-g_{\phi\phi}'g_{t\phi}'h_{t\phi}-g_{\phi\phi}'g_{\phi t}'h_{\phi t}+g_{\phi\phi}'g_{\phi\phi}'h_{tt}\nn\\
\gamma_{t\phi}&=&g_{\phi t}'g_{tt}'h_{\phi\phi}-g_{\phi\phi}'g_{tt}'h_{t\phi}-g_{t\phi}'g_{\phi t}'h_{\phi t}+g_{\phi\phi}'g_{t\phi}'h_{tt}\nn\\
\gamma_{\phi t}&=&g_{\phi t}'g_{tt}'h_{\phi\phi}-g_{\phi\phi}'g_{tt}'h_{t\phi}-g_{\phi t}'g_{t\phi}'h_{\phi t}+g_{\phi t}'g_{\phi\phi}'h_{tt}\nn\\
\gamma_{tt}&=&g_{tt}'g_{tt}'h_{\phi\phi}-g_{tt}'g_{t\phi}'h_{t\phi}-g_{tt}'g_{\phi
t}'h_{\phi t}+g_{t\phi}'g_{\phi t}'h_{tt}\nn \eea \bea
(I)_{\phi\phi}&=&\frac{\sqrt{-g}(\tilde{\epsilon}^{rt\phi})^2}{W}[-\frac{1}{2}(\partial_r\frac{g_{\phi\phi}'}{\sqrt{-g}})\partial_{\phi}(\partial_th_{\phi t}-\partial_{\phi}h_{tt})+\frac{1}{2}(\partial_r\frac{g_{\phi t}'}{\sqrt{-g}})\partial_{\phi}(\partial_th_{\phi\phi}-\partial_{\phi}h_{t\phi})]\nn\\
(I)_{t\phi}&=&\frac{\sqrt{-g}(\tilde{\epsilon}^{rt\phi})^2}{W}[-\frac{1}{2}(\partial_r\frac{g_{\phi\phi}'}{\sqrt{-g}})\partial_t(\partial_th_{\phi t}-\partial_{\phi}h_{tt})+\frac{1}{2}(\partial_r\frac{g_{\phi t}'}{\sqrt{-g}})\partial_t(\partial_th_{\phi\phi}-\partial_{\phi}h_{t\phi})]\nn\\
(I)_{\phi t}&=&\frac{\sqrt{-g}(\tilde{\epsilon}^{rt\phi})^2}{W}[-\frac{1}{2}(\partial_r\frac{g_{tt}'}{\sqrt{-g}})\partial_{\phi}(\partial_{\phi}h_{t\phi }-\partial_{t}h_{\phi\phi})+\frac{1}{2}(\partial_r\frac{g_{t\phi }'}{\sqrt{-g}})\partial_{\phi}(\partial_{\phi}h_{tt}-\partial_{t}h_{\phi t})]\nn\\
(I)_{tt}&=&\frac{\sqrt{-g}(\tilde{\epsilon}^{rt\phi})^2}{W}[-\frac{1}{2}(\partial_r\frac{g_{tt}'}{\sqrt{-g}})\partial_{t}(\partial_{\phi}h_{t\phi
}-\partial_{t}h_{\phi\phi})+\frac{1}{2}(\partial_r\frac{g_{t\phi
}'}{\sqrt{-g}})\partial_{t}(\partial_{\phi}h_{tt}-\partial_{t}h_{\phi
t})]\nn \eea \bea
(II)_{\phi\phi}&=&\frac{\sqrt{-g}}{2}(\partial_r\frac{g_{\phi t}'}{\sqrt{-g}})(-g_{\phi\phi}h_{\phi t}+g_{t\phi}h_{\phi\phi})-\frac{\sqrt{-g}}{2}(\partial_r\frac{g_{\phi\phi}'}{\sqrt{-g}})(g_{t\phi}h_{t\phi}-g_{\phi\phi}h_{tt})\nn\\
(II)_{t\phi}&=&\frac{\sqrt{-g}}{2}(\partial_r\frac{g_{\phi\phi}'}{\sqrt{-g}})(-g_{t\phi}h_{t t}+g_{tt}h_{t\phi})-\frac{\sqrt{-g}}{2}(\partial_r\frac{g_{t\phi}'}{\sqrt{-g}})(g_{t\phi}h_{\phi t}-g_{tt}h_{\phi\phi})\nn\\
(II)_{\phi t}&=&\frac{\sqrt{-g}}{2}(\partial_r\frac{g_{tt}'}{\sqrt{-g}})(-g_{\phi t}h_{\phi \phi}+g_{\phi\phi}h_{\phi t})-\frac{\sqrt{-g}}{2}(\partial_r\frac{g_{\phi t}'}{\sqrt{-g}})(g_{t\phi}h_{t\phi}-g_{\phi\phi}h_{tt})\nn\\
(II)_{tt}&=&\frac{\sqrt{-g}}{2}(\partial_r\frac{g_{t\phi}'}{\sqrt{-g}})(-g_{tt}h_{t\phi
}+g_{\phi
t}h_{tt})-\frac{\sqrt{-g}}{2}(\partial_r\frac{g_{tt}'}{\sqrt{-g}})(g_{\phi
t}h_{t\phi}-g_{tt}h_{\phi\phi})\nn \eea and $\beta$ is defined as
\be \beta=g_{\phi t}'g_{t\phi}'-g_{tt}'g_{\phi\phi}'.  \ee The above
equations are our main results for the rank 2 tensor perturbations
in three-dimensional black hole backgrounds satisfying  the
conditions  (\ref{metriccon}).

For the BTZ black hole, we replace $r$ to $r^2$ and $\partial_r$ to
$\partial_{r^2}$, then we find \bea
(I)_{ij}=(II)_{ij}=0,\hs{3ex}\beta=1,\hs{3ex}\sqrt{-g}=\frac{1}{2}\nn\\
\gamma_{\phi\phi}=h_{tt},\hs{3ex}\gamma_{t\phi}=h_{t\phi},\hs{3ex}\gamma_{\phi t}=h_{\phi t},\hs{3ex}\gamma_{rr}=h_{\phi\phi}\\
\alpha_{\phi\phi}=-2h_{t\phi},\hs{3ex}\alpha_{t\phi}=-h_{\phi\phi}-h_{tt},\hs{3ex}\alpha_{\phi
t}=h_{tt}+h_{\phi\phi},\hs{3ex}\alpha_{tt}=2h_{\phi t}\nn \eea such
that \bea
\tilde{\Delta}h_{tt}&=&m^2h_{tt}+2mh_{\phi t}+2mh_{t\phi}+h_{tt}+2h_{\phi\phi},\nn\\
\tilde{\Delta}h_{t\phi}&=&m^2h_{t\phi}+2mh_{\phi\phi}+2mh_{tt}+h_{t\phi}+2h_{\phi t},\nn\\
\tilde{\Delta}h_{\phi t}&=&m^2h_{\phi t}+2mh_{\phi\phi}+2mh_{tt}+h_{\phi t}+2h_{t\phi},\nn\\
\tilde{\Delta}h_{\phi\phi}&=&m^2h_{\phi\phi}+2mh_{t\phi}+2mh_{\phi
t}+h_{\phi\phi}+2h_{tt}.\nn \eea

For the warped spacetimes, the discussion is similar but more
tedious. It turns out that for the spacelike stretched AdS$_3$ black
hole, the component $h_{tt}$ obeys the equation \be
\tilde{\Delta}h_{tt}=(m^2+4m\nu+3\nu^2)h_{tt}, \ee while for the
self-dual AdS$_3$ black hole, the component $h_{\phi\phi}$ obeys \be
\tilde{\Delta}h_{\phi\phi}=(m^2-4m\nu+3\nu^2)h_{\phi\phi}. \ee


\begin{thebibliography}{}
\bibitem{Andy2010}
 A.~Castro, A.~Maloney and A.~Strominger,
  ``Hidden Conformal Symmetry of the Kerr Black Hole,''
  arXiv:1004.0996 [hep-th].

 \bibitem{Chen:2010xu}
  B.~Chen and J.~Long,
  ``Real-time Correlators and Hidden Conformal Symmetry in Kerr/CFT
  Correspondence,'' JHEP {\bf 1006}, 018 (2010)
  [arXiv:1004.5039 [hep-th]].

\bibitem{Chen:2010bh}
  B.~Chen and J.~Long,
  ``On Holographic description of the Kerr-Newman-AdS-dS black holes,''JHEP {\bf 1008}, 065 (2010),   arXiv:1006.0157 [hep-th].

\bibitem{Chen:2010fr}
  B.~Chen, J.~Long and J.~j.~Zhang,
  ``Hidden Conformal Symmetry of Extremal Black Holes,''
  arXiv:1007.4269 [hep-th].


\bibitem{71}C. Krishnan, JHEP 1007 (2010) 039, [arXiv:1004.3537[hep-th]];
[arXiv:1005.1629[hep-th]]. C.-M. Chen and J.-R. Sun,
[arXiv:1004.3963[hep-th]]. Y.-Q. Wang and Y.-X. Liu,
[arXiv:1004.4661[hep-th]]. R. Li, M.-F. Li and J.-R. Ren,
[arXiv:1004.5335[hep-th]].D. Chen, P. Wang and H. Wu,
[arXiv:1005.1404[gr-qc]]. M. Becker, S. Cremonini and W. Schulgin,
[arXiv:1005.3571[hep-th]]. H. Wang, D. Chen, B. Mu and H. Wu,
[arXiv:1006.0439[gr-qc]]. C.-M. Chen, Y.-M. Huang, J.-R. Sun, M.-F.
Wu and S.-J. Zou, [arXiv:1006.4092[hep-th]];
[arXiv:1006.4097[hep-th]]. Y.~Matsuo, T.~Tsukioka and C.~M.~Yoo,
    arXiv:1007.3634 [hep-th].K.~N.~Shao and Z.~Zhang,
    arXiv:1008.0585 [hep-th].
M. R. Setare, V. Kamali, [arXiv:1008.1123[hep-th]]. A. M.
Ghezelbash, V. Kamali, M. R. Setare, arXiv:1008.2189v1 [hep-th].



\bibitem{quasinomral1999}
H.P. Nollert, Class.Quant.Grav.16:R159-R216,(1999). K.~D.~Kokkotas
and B.~G.~Schmidt,
    Living Rev.\ Rel.\  {\bf 2}, 2 (1999)
  [arXiv:gr-qc/9909058]. E.~Berti, V.~Cardoso and A.~O.~Starinets,
  Class.\ Quant.\ Grav.\  {\bf 26}, 163001 (2009)
  [arXiv:0905.2975 [gr-qc]].



\bibitem{Horowitz:1999jd}
  G.~T.~Horowitz and V.~E.~Hubeny,
  ``Quasinormal modes of AdS black holes and the approach to thermal
  equilibrium,''
  Phys.\ Rev.\  D {\bf 62}, 024027 (2000)
  [arXiv:hep-th/9909056].


\bibitem{Briminghan01}
D.~Birmingham, I.~Sachs and S.~N.~Solodukhin, ``Conformal Field
Theory Interpretation of Black Hole Quasinomral Modes''
 Phys.\ Rev.\ Lett.\  {\bf 88}:151301,(2002)
 [hep-th/0112055].


\bibitem{Son:2002sd}
  D.~T.~Son and A.~O.~Starinets,
  ``Minkowski-space correlators in AdS/CFT correspondence: Recipe and
  applications,''
  JHEP {\bf 0209}, 042 (2002)
  [arXiv:hep-th/0205051].

\bibitem{Cardy:1984rp}
  J.~L.~Cardy,
  ``Conformal invariance and universality in finite-size scaling,''
  J.\ Phys.\ A {\bf 17}, L385 (1984).

\bibitem{Sachs08}
Ivo Sachs and Sergey N. Solodukhin, ``Quasinormal Modes in
Topologically Massive Gravity'' JHEP 0808:003,(2008)
[hep-th/0806.1788].

\bibitem{Afshar:2010ii}
 H.~R.~Afshar, M.~Alishahiha and A.~E.~Mosaffa,
 ``Quasinormal Modes of Extremal BTZ Black Holes in TMG,''
 JHEP {\bf 1008}, 081 (2010)
 [arXiv:1006.4468 [hep-th]].

\bibitem{BTZ}
M$\acute{a}$ximo Ba$\tilde{n}$ados , Claudio Teitelboim and Jorge
Zanelli, ``The Black hole in three-dimensional space-time''
Phys.Rev.Lett.69:1849-1851,(1992) [hep-th/9204099].

\bibitem{Cardoso:2001hn}
  V.~Cardoso and J.~P.~S.~Lemos,
  ``Scalar, electromagnetic and Weyl perturbations of BTZ black holes:  Quasi
  normal modes,''
  Phys.\ Rev.\  D {\bf 63}, 124015 (2001)
  [arXiv:gr-qc/0101052].



\bibitem{Andy08}D. Anninos, W. Li, M. Padi, W. Song and A.
Strominger, {\it Warped $AdS_3$ black holes}, [arXiv:0807.3040].

\bibitem{Anninos:2009zi}
  D.~Anninos, M.~Esole and M.~Guica,
  ``Stability of warped AdS3 vacua of topologically massive gravity,''
  JHEP {\bf 0910}, 083 (2009)
  [arXiv:0905.2612 [hep-th]].


\bibitem{ChenXu09}B.~Chen and Z.~b.~Xu,
  ``Quasinormal modes of warped $AdS_3$ black holes and AdS/CFT
  correspondence,'', Phys. Lett. B 675(2009)246-251.
  arXiv:0901.3588 [hep-th].

\bibitem{ChenXu2}B.~Chen and Z.~b.~Xu,
  ``Quasinormal modes of warped black holes and warped AdS/CFT
  correspondence,'' JHEP {\bf 11}(2009)091,
  arXiv:0908.0057 [hep-th].

\bibitem{Fareghbal:2010yd}
  R.~Fareghbal,
  ``Hidden Conformal Symmetry of Warped AdS$_3$ Black Holes,''
  arXiv:1006.4034 [hep-th].

\bibitem{Bin10selfdual}
Bin Chen, George Moutsopoulos and Bo Ning,
``Self-Dual Warped AdS3 Black Holes''
arXiv:1005.4175.

\bibitem{Li:2010zr}
  R.~Li, M.~F.~Li and J.~R.~Ren,
  ``Hidden Conformal Symmetry of Self-Dual Warped AdS$_3$ Black Holes in
  Topological Massive Gravity,''
  arXiv:1007.1357 [hep-th].

\bibitem{Li:2010sv}
  R.~Li and J.~R.~Ren,
  ``Quasinormal Modes of Self-Dual Warped AdS$_3$ Black Hole in Topological
  Massive Gravity,''
  arXiv:1008.3239 [hep-th].


\bibitem{Hartman:2009nz}
  T.~Hartman, W.~Song and A.~Strominger,
  ``Holographic Derivation of Kerr-Newman Scattering Amplitudes for General
  Charge and Spin,''
  JHEP {\bf 1003}, 118 (2010)
  [arXiv:0908.3909 [hep-th]].

\bibitem{Chen:2010ni}
  B.~Chen and C.~S.~Chu,
  ``Real-time correlators in Kerr/CFT correspondence,''
  JHEP {\bf 1005}, 004 (2010)
  [arXiv:1001.3208 [hep-th]].


\end{thebibliography}
\end{document}